\documentclass[11pt]{article}    

\usepackage{amsthm}
\usepackage{amsmath}
\usepackage{amssymb}
\usepackage[small]{caption}
\usepackage{epsfig}

\newtheorem{theorem}{Theorem}
\newtheorem{lemma}{Lemma}

\def\L{\mathcal L}

\newcommand{\old}[1]{{}}
\newcommand{\later}[1]{{}}

\def\etal{{et~al.}}
\def\eg{{e.g.}}
\def\ie{{i.e.}}

\newcommand{\area}{{\rm Area}}
\newcommand{\conv}{{\rm conv}}
\newcommand{\per}{{\rm per}}
\newcommand{\arc}{{\rm arc}}
\newcommand{\len}{{\rm len}}

\newcommand{\opt}{\mathrm{OPT}}
\newcommand{\dist}{\mathrm{dist}}

\newcommand{\RR}{\mathbb{R}}

\def\eps{\varepsilon}
\providecommand{\intd}[0]%
{\;\mbox{d}}

\title{Opaque sets\footnote{A preliminary version
of this paper appeared in 
\emph{Proceedings of the 14th International Workshop on
Approximation Algorithms for Combinatorial Optimization Problems}
(APPROX 2011), Princeton, New Jersey, August 2011, LNCS Vol. 6845,
pp.~194--205.}}

\author{Adrian Dumitrescu\thanks{Department of Computer Science,
University of Wisconsin--Milwaukee, USA\@.
Email:~\texttt{dumitres@uwm.edu}.
Supported in part by NSF CAREER grant CCF-0444188
and by NSF grant DMS-1001667.
Part of the research by this author was done at 
Ecole Polytechnique F\'ed\'erale de Lausanne.}
\and
Minghui Jiang\thanks{%
Department of Computer Science,
Utah State University,
Logan, USA\@.
Email: \texttt{mjiang@cc.usu.edu}.
Supported in part by NSF grant DBI-0743670.}
\and
J\'anos Pach\thanks{Ecole Polytechnique F\'ed\'erale de Lausanne and
  City College, New York. Email:~\texttt{pach@cims.nyu.edu}.
Research partially supported by NSF grant CCF-08-30272,
grants from OTKA, SNF, and PSC-CUNY.
}}

\begin{document}

\maketitle

\begin{abstract}
The problem of finding ``small'' sets that meet every straight-line
which intersects a given convex region was initiated by Mazurkiewicz in 1916.
We call such a set an \emph{opaque set} or a \emph{barrier} for that region.
We consider the problem of computing the shortest barrier for a given
convex polygon with $n$ vertices. No exact algorithm is currently known
even for the simplest instances such as a square or an equilateral triangle. 
For general barriers, we present an approximation algorithm with ratio
$\frac{1}{2}+ \frac{2 +\sqrt{2}}{\pi}=1.5867\ldots$.
For connected barriers we achieve the approximation ratio $1.5716$,
while for single-arc barriers we achieve the approximation ratio
$\frac{\pi+5}{\pi+2} = 1.5834\ldots$. All three algorithms run in $O(n)$ time. 
We also show that if the barrier is restricted to the 
(interior and the boundary of the)
input polygon, then the problem admits a fully
polynomial-time approximation scheme for the connected case and a
quadratic-time exact algorithm for the single-arc case.

\end{abstract}

\section{Introduction} \label{sec:intro}

The problem of finding small sets that block every line passing
through a unit square was first considered by Mazurkiewicz in
1916~\cite{Ma16}; see also~\cite{Ba59,GM55}. 
Let $C$ be a convex body in the plane. Following Bagemihl~\cite{Ba59},
we call a set $B$ an \emph{opaque set} or a \emph{barrier} for $C$, if
it meets all lines that intersect $C$.
A barrier may consist of one or
more rectifiable arcs. It does not need to be connected and its
portions may lie anywhere in the plane, including the exterior of $C$;
see~\cite{Ba59,Br92}. 
We restrict our attention to barriers for convex bodies because
every line that intersects a non-convex object must also intersect its
convex hull. 

\emph{What is the length of the shortest barrier for a given convex
body $C$?} In spite of considerable efforts, the answer to this
question is not known even for the simplest instances of $C$,
such as a square, a disk, or an equilateral triangle; see~\cite{Cr69},
~\cite[Problem~A30]{CFG91},~\cite{E82},~\cite{FM86},~\cite{FMP84},
~\cite[Section~8.11]{F03},~\cite[Problem~12]{H78}.
The three-dimensional analogue of this problem was raised by Martin
Gardner~\cite{G90}; see also~\cite{AG08,Br92}. 

\begin{figure}[htbp]
\centerline{\epsfxsize=6.3in \epsffile{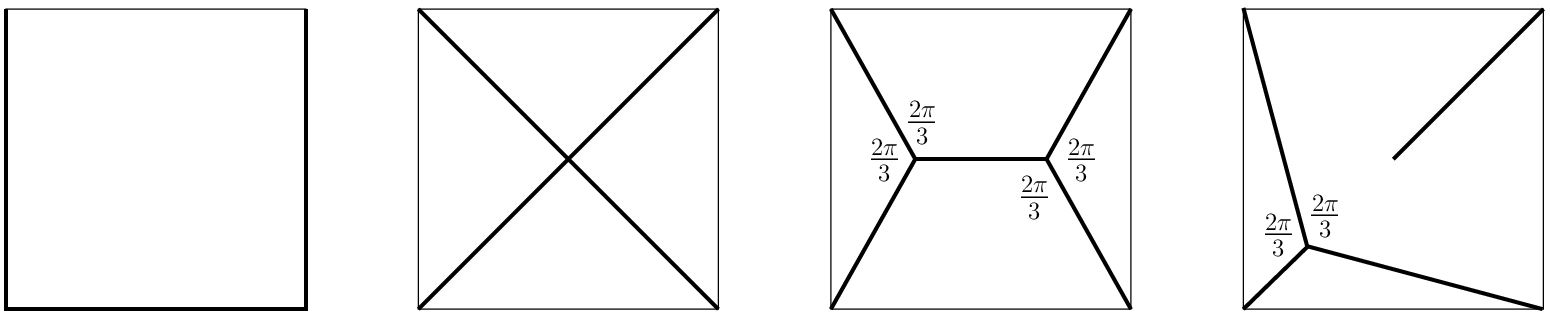}}
\caption{Four barriers for the unit square.  
From left to right: 1: single-arc; 2--3: connected; 4: disconnected.
The first three from the left have lengths $3$,
$2\sqrt{2}=2.8284\ldots$, and $1 +\sqrt{3}= 2.7320\ldots$. 
Right: The diagonal segment $[(1/2,1/2),(1,1)]$ together with
  three segments connecting the corners $(0,1)$, $(0,0)$, $(1,0)$ to
  the point $(\frac{1}{2}-\frac{\sqrt3}{6},\frac{1}{2}-\frac{\sqrt3}{6})$
  yield a barrier of length $\sqrt2 + \frac{\sqrt6}{2}= 2.6389\ldots$.}
\label{f1}
\end{figure}

A barrier blocks any line of sight across the region $C$ or detects
any ray that passes through it. Motivated by potential applications in
guarding and surveillance, the problem of short barriers has been
studied by several research communities. Recently, it circulated in
internal publications at the Lawrence Livermore National Laboratory~\cite{DO08}. 
The shortest barrier known for the square,  of length $2.6389\ldots$,
is illustrated in Fig.~\ref{f1}~(right). It is conjectured to be
optimal. The current best lower bound is $2$, established by Jones~\cite{J64}.

Another real-world application is mentioned by Faber~\etal~\cite{FMP84,FM86}: 
A repairman from a telephone company, while repairing buried cable,
has discovered that often the cable is not directly under the marker
which is supposed to be erected above it.
Assuming that the cable is straight and is always within 2 meters
from the marker in a horizontal plane, what is shortest length of
a trench that the repairmen has to dig such that the cable is guaranteed
to be found? In the terminology of the opaque set problem,
the disk of radius 2 meters centered at the marker is the convex body,
the possible locations of the cable are the lines intersecting
the convex body, and the trench is the barrier.

Some entertaining variants of the opaque set problem appeared in different
forms~\cite{J80,K86,K87}; see also~\cite[Problem~A30]{CFG91}. 
For instance, what should a swimmer at sea do in a thick fog
if he knows that he is within a mile of a straight shoreline?
Here the convex body is the disk of radius one mile centered at the start
location of the swimmer, and the barrier is the route taken by the swimmer.
This is almost the same problem as that for the telephone company except that
the barrier here is restricted to be a single curve originating from
the disk center.

\paragraph{Related work.}
The type of curve barriers considered may vary: the most restricted are
barriers made from single continuous arcs, then connected barriers,
and lastly, arbitrary (possibly disconnected) barriers. For the unit
square, the shortest known in these three categories have lengths $3$, 
$1 +\sqrt{3}= 2.7320\ldots$ and $\sqrt2 + \frac{\sqrt6}{2}=
2.6389\ldots$, respectively. They are depicted in Fig.~\ref{f1}. 
Interestingly, it has been shown by Kawohl~\cite{K00} that the barrier in
Fig.~\ref{f1}~(right) is optimal in the class of curves with at most
two components (there seems to be an additional implicit assumption
that the barrier is restricted to the interior of the square).
For the unit disk, the shortest known barrier consists
of three arcs. See also~\cite{FM86,F03}.

If instead of curve barriers, we want to find \emph{discrete} barriers
consisting of as few points as possible with 
the property that every line intersecting $C$ gets closer than
$\eps>0$ to at least one of them in some fixed norm, we arrive
at a problem raised by L\'aszl\'o Fejes T\'oth~\cite{FT73,FT74}. 
The problem has been later coined suggestively 
as the ``point goalie problem''~\cite{RS03}. For instance, if $C$ is
an axis-parallel unit square, and we consider the \emph{maximum norm},
the problem was studied by B\'ar\'any and F\"{u}redi~\cite{BF85}, Kern
and Wanka~\cite{KW90}, Valtr~\cite{V94}, and Richardson and Shepp~\cite{RS03}. 
Makai and Pach~\cite{MP83} considered another variant of the  question,
in which we have a larger class of functions to block. 

The problem of short barriers has attracted many other researchers and has
been studied at length; see also~\cite{Cr69,EP80,H78,KKNX09,M80}. 
Obtaining lower bounds for many of these problems appears to be
notoriously hard.  For instance in the point goalie problem for the
unit disk (with the Euclidean norm), while the trivial lower bound is
$1/\eps$, as given by the opaqueness condition in any one direction,
the best lower bound known is only $1.001/\eps$ as established in
\cite{RS03} via a complicated proof.

\paragraph{Our results.}

Even though we have so little control on the shape or length of optimal
barriers, for any convex polygon, barriers whose lengths are
somewhat longer can computed efficiently. 
Let $P$ be a given convex polygon with $n$ vertices.

\begin{enumerate}
\item A (possibly disconnected) barrier for $P$,
whose length is at most $\frac{1}{2}+ \frac{2 +\sqrt{2}}{\pi}=1.5867\ldots$
times the optimal, can be computed in $O(n)$ time.  

\item A connected polygonal barrier  whose length is at most 
$1.5716$ times the optimal can be computed in $O(n)$ time.  

\item A single-arc polygonal barrier whose length is at most 
$\frac{\pi+5}{\pi+2} = 1.5834\ldots$ times the optimal can be 
computed in $O(n)$ time.   

\item For interior single-arc barriers we present an algorithm  that finds
an optimal barrier in $O(n^2)$ time.

\item For interior connected barriers we present an algorithm  that finds
a barrier whose length is at most $(1+\eps)$ times the optimal in
polynomial time.  

\end{enumerate}

It might be worth mentioning to avoid any confusion: the approximation
ratios are for each barrier class, that is, the length of the barrier
computed is compared to the optimal length in the corresponding class;
and of course these optimal lengths might differ. For instance 
the connected barrier computed by the approximation algorithm with ratio 
$1.5716$ is \emph{not} necessarily shorter
than the (possibly disconnected) barrier computed by the 
approximation algorithm with the larger ratio 
$\frac{1}{2}+ \frac{2 +\sqrt{2}}{\pi}=1.5867\ldots$.

However, we believe that the approximation ratios of the first two
algorithms mentioned above are substantially better than $1.57$. 
In support of this belief, we present a couple of lower bound examples 
for which the ratios are below $1.1$.

\section{Preliminaries} \label{sec:prelim}

\paragraph{Definitions and notations.}

For a curve $\gamma$, let $|\gamma|$ denote the length of
$\gamma$. Similarly, if $\Gamma$ is a set of curves, let $|\Gamma|$ denote
the total length of the curves in $\Gamma$.
When there is no danger of confusion, $|A|$ also denotes the
cardinality of a set $A$. 

In order to be able to speak of the \emph{length}
$\len(B)$ of a barrier $B$, we restrict our attention to
rectifiable barriers. A \emph{rectifiable curve} is a curve of finite length.
A \emph{rectifiable barrier} is the union of a countable set of
\emph{rectifiable curves}, $\Gamma=\cup_{i=1}^\infty \gamma_i$, 
where $\sum_{i=1}^\infty |\gamma_i| <\infty$ 
(or $\Gamma=\cup_{i=1}^n \gamma_i$ for some $n$).
A~\emph{segment barrier} is a barrier consisting of straight-line
segments (or polygonal paths). A curve is a \emph{convex curve} if it
is a subset of the boundary of a convex set. 

We first show that the shortest segment barrier is
not much longer than the shortest rectifiable one. 

\begin{lemma}\label{L1}
Let $B$ be a rectifiable barrier for a convex body $C$ in the plane.
Then, for any $\eps>0$, there exists a segment barrier $B_{\eps}$ for $C$,
consisting of a countable set of straight-line segments, such that
$\len(B_{\eps}) \leq (1+\eps)\, \len(B)$.
\end{lemma}
\begin{proof}
Suppose that $B$ is the union of a countable set of rectifiable curves.
Decompose each rectifiable curve in the set into a sequence of convex curves
by cutting at points where the curvature changes sign or the curve crosses
itself.
Then $B$ becomes the union of a countable set $\Gamma$ of convex curves.

For each convex curve $\gamma_i \in \Gamma$,
let $C_i$ be the convex hull of $\gamma_i$,
and let $B'_i$ be an arbitrary barrier for $C_i$.
Note that every line intersecting $C$ is blocked by some curve $\gamma_i$,
every line blocked by $\gamma_i$ intersects $C_i$,
and every line intersecting $C_i$ is blocked by $B'_i$.
Thus the union $B'$ of the barriers $B'_i$ for $C_i$ is a barrier for $C$.
Since any convex curve $\gamma: [0,1]\rightarrow\RR^2$ 
is a barrier for its convex hull $\conv(\gamma)$,
it suffices to prove the lemma for barriers $B$ consisting of a single 
convex curve $\gamma$, with $C=\conv(\gamma)$.

In this simple case, we can approximate the convex curve $\gamma$
by a polygonal path $\gamma'$ with the same endpoints,
which avoids the interior of $C$,
such that $|\gamma'| \leq (1+\eps) \, |\gamma|$.
Then the union of the segments in $\gamma'$ is the desired segment barrier
$B_{\eps}$ for $C$.
\end{proof}

Denote by $\per(C)$ the perimeter of a convex body $C$ in the plane.
The following lemma providing a lower bound on the length of an optimal
barrier for $C$ in terms of $\per(C)$, is used in the analysis of
our approximation algorithms. Its proof is folklore; see for
instance~\cite{FMP84}.

\begin{lemma}\label{L2}
Let $C$ be a convex body in the plane and let $B$ be a barrier for $C$.
Then the length of $B$ is at least $\frac12 \cdot \per(C)$.
\end{lemma}
\begin{proof} 
Let
$B=\{s_1,\ldots,s_n\}$ consist of $n$ segments of lengths $\ell_i=|s_i|$, where
$L=|B|=\sum_{i=1}^n \ell_i$.
Let $\alpha_i \in [0,\pi)$ be the angle made by $s_i$ with
the $x$-axis. For each direction $\alpha \in [0,\pi)$, the
blocking (opaqueness) condition for a convex body $C$
requires
\begin{equation} \label{E1}
\sum_{i=1}^n \ell_i |\cos(\alpha-\alpha_i)| \geq W(\alpha).
\end{equation}
Here $W(\alpha)$ is the width of $C$ in direction $\alpha$, \ie, the
minimum width of a strip of parallel lines enclosing $C$, whose lines
are orthogonal to direction $\alpha$. By integrating this inequality
over the interval $[0,\pi]$, one gets: 
\begin{equation} \label{E2}
\sum_{i=1}^n \ell_i \int_0^{\pi} |\cos(\alpha-\alpha_i)| \intd \alpha
\geq \int_0^{\pi} W(\alpha) \intd \alpha.
\end{equation}
According to Cauchy's surface area formula~\cite[pp.~283--284]{PA95},
for any planar convex body $C$, we have
\begin{equation} \label{E3}
\int_0^{\pi} W(\alpha) \intd \alpha = \per(C).
\end{equation}
Since
$$ \int_0^{\pi} |\cos(\alpha-\alpha_i)| \intd \alpha =2, $$
we get
\begin{equation} \label{E4}
2L= \sum_{i=1}^n 2\ell_i \geq \per(C) \ \Rightarrow \
L \geq \frac12 \cdot \per(C),
\end{equation}
as required.
\end{proof}

For instance, for the square, $\per(C)=4$, and Lemma~\ref{L2}
immediately gives $L \geq 2$, the lower bound of Jones~\cite{J64}). 

\paragraph{Remark.} Obviously, the boundary of $C$, $\partial C$, is a
barrier for $C$ of length $\per(C)$. Consequently, once Lemma~\ref{L2}
is established, a $2$-approximation (for each type of barrier) follows
immediately. 
A much better approximation can be obtained
for ``thin'' convex bodies
whose widths are much smaller than their diameters (and hence much smaller
than their perimeters).
For a convex body of width $w$ and perimeter $p$,
algorithm {\bf A1} in Section~\ref{sec:connected}
constructs a single-arc barrier of length $p/2 + w$,
which is close to the lower bound $p/2$ when $w$ is relatively small.
This also shows that the lower bound in Lemma~\ref{L2} is almost tight
for thin convex bodies.

\medskip
A key fact in the analysis of our approximation algorithms is the
following lemma. This inequality is implicit in~\cite{W93}; another
proof can be found in~\cite{DJ12}.  

\begin{lemma}\label{L10}
Let $P$ be a convex polygon. Then the minimum-perimeter rectangle $R$
containing $P$ satisfies $\per(R) \leq \frac{4}{\pi}\,\per(P)$.
\end{lemma}

Let $P$ be a convex polygon with $n$ vertices.
Let $\opt_{\rm arb}(P)$, $\opt_{\rm conn}(P)$ and $\opt_{\rm arc}(P)$
denote optimal barrier lengths of the types arbitrary, connected, and
single-arc. Observe the following inequalities: 
\begin{equation} \label{E5}
\opt_{\rm arb}(P) \leq \opt_{\rm conn}(P) \leq \opt_{\rm arc}(P).
\end{equation}

We first deal with connected barriers, and then with arbitrary (\ie,
possibly disconnected) barriers.

\section{Connected barriers} \label{sec:connected}

\begin{theorem} \label{T1}
Given a convex polygon $P$ with $n$ vertices, a connected polygonal
barrier  whose length is at most $1.5716$ times longer than the
optimal can be computed in $O(n)$ time.  
\end{theorem} 
\begin{proof}
We start with the following algorithm {\bf A1} that computes a connected barrier
consisting of a single-arc; refer to Fig.~\ref{f4}.
\begin{figure}[htbp]
\centerline{\epsfxsize=3in \epsffile{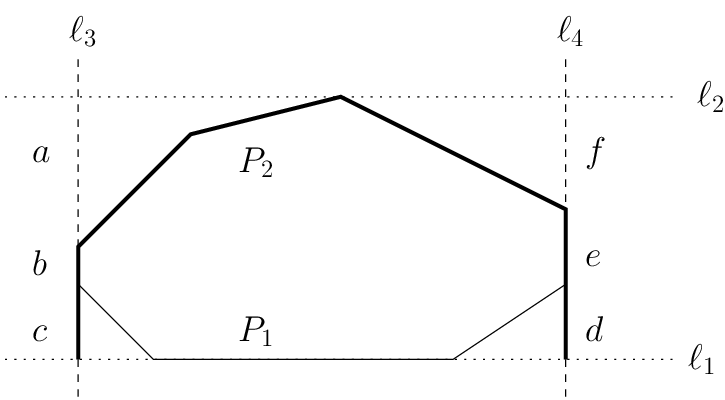}}
\caption{The approximation algorithm {\bf A1} returns $B_2$ (in bold lines).}
\label{f4}
\end{figure}
First compute a parallel strip of minimum width enclosing $P$.
Assume w.l.o.g.\ that the strip is bounded by the two
horizontal lines $\ell_1$ and $\ell_2$.
Second, compute a minimal orthogonal 
(\ie, vertical) strip enclosing $P$, bounded by the
two vertical lines $\ell_3$ and $\ell_4$. Let $a,b,c,d,e,f$ be the six
segments on $\ell_3$ and $\ell_4$ as shown in the figure; here $b$ and
$e$ are the two (possibly degenerate) segments on the boundary of $P$. 
Let $P_1$ be the polygonal path (on $P$'s
boundary) between the lower vertices of $b$ and $e$.
Let $P_2$ be the polygonal path (on $P$'s
boundary) between the top vertices of $b$ and $e$.

Consider the following two barriers for $P$:
$B_1$ consists of the polygonal path $P_1$ extended upward at both
ends until they reach $\ell_2$.
$B_2$ consists of the polygonal path $P_2$ extended downwards at both
ends until they reach $\ell_1$. The algorithm returns the shorter of
the two. We show below that its approximation ratio is at most 
$\frac{\pi+5}{\pi+2} = 1.5834\ldots$.

Let $p$, $w$, and $r$, respectively,
be the perimeter, the width, and the in-radius of $P$.
Clearly
$$ |P_1| + |P_2| + |b| + |e|= p. $$
We have the following equalities:
\begin{align*} 
|B_1| &= |a| + |b| +|P_1| + |e| + |f|,\\
|B_2| &= |c| + |b| +|P_2| + |e| +|d|.
\end{align*}
By adding them up we get
$$ |B_1| + |B_2| = |P_1| + |P_2| + |b| + |e| + 2w  = p +2w. $$
Hence
$$ \min\{ |B_1|, |B_2| \} \leq p/2 + w. $$

By Blaschke's Theorem~\cite{Bl56} (see also~\cite[Exercise 2-5]{YB61}), 
every planar convex body of width $w$ contains a disk of radius
$w/3$, hence $r \ge w/3$. This inequality cannot be improved: equality
is attained for the equilateral triangle.
According to a result of Eggleston~\cite{E82},
the optimal connected barrier for a disk of radius $r$
has length $(\pi + 2)r$.
It follows that the optimal connected barrier for $P$
has length at least $(\pi + 2) w/3$.
By Lemma~\ref{L2}, $p/2$ is another lower bound on the optimal solution.
Thus the approximation ratio of the algorithm {\bf A1} is at most
\begin{align*}
\frac{p/2 + w}{\max\{ (\pi + 2) w/3,\, p/2 \}} &=
\min\left\{
	\frac{p/2 + w}{(\pi + 2) w/3},\,
	\frac{p/2 + w}{p/2}
\right\}\\
&= \min\left\{
	\frac{3}{2(\pi + 2)}\cdot \frac{p}{w} + \frac{3}{\pi + 2},\,
	1 + 2\cdot \frac{w}{p}
\right\}.
\end{align*}

One can check that the quadratic equation
$$
	\frac{3x}{2(\pi + 2)} + \frac{3}{\pi + 2} =
	1 + \frac{2}{x}
$$
has one positive real root
$$
x_0 = \frac{2(\pi + 2)}{3}.
$$

Consequently, the approximation ratio of the algorithm {\bf A1} is at most
$1 + \frac{3}{\pi + 2} = \frac{\pi + 5}{\pi + 2} = 1.5834\ldots$.
Clearly the algorithm takes $O(n)$ time, since computing the width of
$P$ takes $O(n)$ time~\cite{PS85,T83}, and the two barriers $B_1$ and
$B_2$ can be computed within the same time.

We next achieve a better approximation, $1.5716$, by means of a more
elaborated approach. The idea is to do something different when 
$P$ is ``close to'' an equilateral triangle. In this case, one of the
two barriers $B_1$ and $B_2$ computed by algorithm {\bf A1} is
substantially shorter than the average of the two, namely,
$ \min\{ |B_1|, |B_2| \} $ is substantially shorter than 
$(|B_1| + |B_2|)/2$, and the previous argument becomes wasteful. 
Our revised algorithm is {\bf A2}.

To explain the algorithm, we need to enter the details of the proof of
Blaschke's Theorem, as given in~\cite[Exercise 6-2]{YB61}. 
Let $\Omega$ be a largest circle contained in $P$; 
let $r$ be its radius.  Then $\Omega$ either contains two
diametrically opposite points of $P$, or else it contains three boundary points
of $P$ which form an acute triangle. In the former (easier)
case, $r =w/2$, and this yields a much better approximation than that obtained
earlier using the inequality $r \geq w/3$; to put it short, 
this is not the bottleneck case. Assume therefore that $\Omega$ is
incident to three boundary points, $A,B,C \in P$ which form an acute
triangle, $\Delta{ABC}$. Then the supporting lines at $A$, $B$, $C$ must
form a triangle $T=\Delta{A'B'C'}$ which is circumscribed to both the
polygon $P$ and the circle $\Omega$. Denote the sides of this triangle 
by $a$, $b$, $c$, where $a$ is a longest side, and the corresponding
altitudes by $h_a$, $h_b$, $h_c$. Denote by $w_a$, $w_b$, $w_c$ the
widths of $P$ in the directions of $a$, $b$ and $c$, respectively. 
Obviously, we have $h_a \geq w_a \geq w$, $h_b \geq w_b \geq w$, 
and $h_c \geq w_c \geq w$. See Fig.~\ref{f8}. 
\begin{figure}[htbp]
\centerline{\epsfxsize=3.3in \epsffile{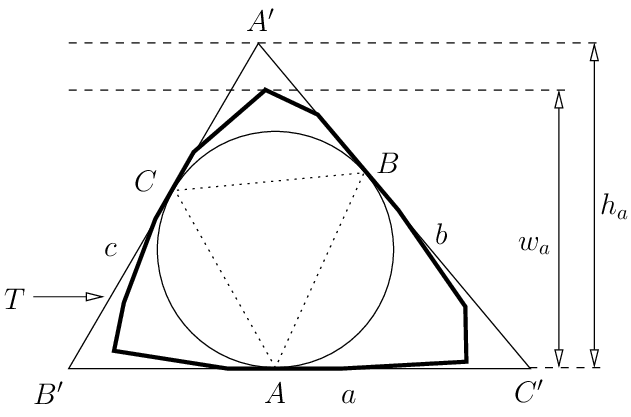}}
\caption{$P$ (in bold lines) and $T$.}
\label{f8}
\end{figure}

We now present the revised algorithm. 
The algorithm {\bf A2} first computes the two barriers $B_1$ and $B_2$
as done by algorithm {\bf A1}. In addition, it also computes a third
barrier, $B_3$, which is a Steiner minimal tree of the three points 
$A'$, $B'$, $C'$, if they exist; otherwise $B_3$ is undefined and
$|B_3|=\infty$. Since $P$ is contained in $T$, $B_3$, which is a
connected barrier for $T$, is also a connected barrier for $P$. 
The algorithm then returns the shorter of the three barriers, $B_1$,
$B_2$, $B_3$.     

Recall that a Steiner minimal tree of three 
points that determine no angle larger or equal to $2\pi/3$ is a star,
whose any two consecutive edges make an angle of $2\pi/3$ between them;
see \eg,~\cite{GP68}, or~\cite[Ch.~6]{RT57}. 

Returning now to the proof of Blaschke's Theorem,
the area of $T$ can written is several ways:
\begin{equation} \label{E6}
\area(T)=\frac{(a+b+c)r}{2}= 
\frac{a \, h_a}{2}= \frac{b \, h_b}{2}= \frac{c \, h_c}{2}.
\end{equation}
Since $a \geq b$, $a \geq c$ it follows that
\begin{equation} \label{E7}
r= \frac{a}{a+b+c} \, h_a \geq \frac{h_a}{3} 
\geq \frac{w_a}{3} \geq \frac{w}{3}.
\end{equation}

This concludes the proof of Blaschke's Theorem. 
Observe that if $a$ is somewhat larger than $(a+b+c)/3$, 
then $r$ is somewhat larger than $w/3$, and one could use this
improved bound to get a better approximation ratio as in the
analysis of Case 1. 
We next analyze the approximation ratio of algorithm {\bf A2}.
We can assume w.l.o.g. that the perimeter of $T$ is $1$, \ie,
$a+b+c=1$, and further that $c \leq b$. Then $a \geq 1/3$. 
We make use of two parameters $\lambda$ and $\delta = 3 - 1 / \lambda$, 
where $1/3 < \lambda \leq 3/8$ and correspondingly
$0 < \delta \leq 1/3$, which will be later set
to $\lambda=0.3403\ldots$ and $\delta=0.0615\ldots$
in order to optimize the approximation ratio of {\bf A2}. 
We distinguish two cases:

\medskip
\emph{Case 1.} $a \geq \lambda$. Then according to~\eqref{E7}, we 
have $r = a h_a \geq a w_a \geq aw \geq \lambda w$. 
Similar to the previous analysis of {\bf A1}, the 
approximation ratio of {\bf A2} is at most
\begin{align*}
\frac{p/2 + w}{\max\{ (\pi + 2) \lambda w,\, p/2 \}}
&=
\min\left\{
	\frac{p/2 + w}{(\pi + 2) \lambda w},\,
	\frac{p/2 + w}{p/2}
\right\} \\
&=
\min\left\{
	\frac{1}{2\lambda(\pi + 2)}\cdot \frac{p}{w} + 
\frac{1}{\lambda(\pi + 2)},\,
	1 + 2\cdot \frac{w}{p}
\right\}.
\end{align*}
As before, one can easily check that the quadratic equation
$$
	\frac{x}{2\lambda(\pi + 2)} + \frac{1}{\lambda(\pi + 2)} =
	1 + \frac{2}{x}
$$
has one positive real root
$$
x_0 = 2\lambda(\pi + 2). 
$$
Under the assumption in Case 1, it follows that the approximation 
ratio is at most 
$$ 1 + \frac{2}{x_0}
= 1 + \frac{1}{\lambda(\pi + 2)}
$$
For future reference, set
\begin{equation} \label{E8}
\rho_1 := 1 + \frac{1}{\lambda(\pi + 2)}
\end{equation}

\medskip
\emph{Case 2.} $a \leq \lambda$. Obviously, we also have $b,c \leq \lambda$.
Then $b= 1-a-c\geq 1-2\lambda$, and similarly, $c \geq 1-2\lambda$. To summarize,
\begin{equation} \label{E9}
1-2\lambda \leq a,b,c \leq \lambda, \ \ a \geq \frac13.
\end{equation}
Recall that $a \leq \lambda \leq 3/8 <\sqrt2 -1$, which implies
that $a^2 < 2(1-a)^2/4 \leq b^2 + c^2$. It follows that $T$ is an
acute triangle. 
We further distinguish two sub-cases, Case 2.1 and Case 2.2. 

\smallskip
\emph{Case 2.1.} At least one of the following three inequalities holds:
(i) $w_a \leq (1-\delta) h_a$; 
(ii) $w_b \leq (1-\delta) h_b$;
(iii) $w_c \leq (1-\delta) h_c$.
Let $\xi \in \{a,b,c\}$ and assume that $w_\xi \leq (1-\delta) h_\xi$. 
Then~\eqref{E6} and~\eqref{E9} yield 
$$
r = \xi h_\xi
\geq (1-2\lambda) \, \frac{w_\xi}{1-\delta}
\geq (1-2\lambda) \, \frac{w}{1/\lambda-2}
= \lambda w.
$$
As in the analysis of Case 1,
it follows that the approximation ratio is again at most $\rho_1$
under the assumption in Case 2.1.

\smallskip
\emph{Case 2.2.} None of the inequalities in Case 2.1 holds. 
We then have $w_\xi \geq (1-\delta) h_\xi$, for each $\xi \in \{a,b,c\}$. 
Construct a triangle $T'$ containing $P$ as described below, and as
shown in Fig.~\ref{f9}. Assume w.l.o.g. that the side $a$ is 
the horizontal base of $T$. 
Consider the three lines, $\ell_a$, $\ell_b$, $\ell_c$, 
each parallel to the corresponding side of $T$:
a line $\ell_a$ parallel to $a$ and tangent to $P$ from above, etc.
Observe that the triangles $T$ and $T'$ are similar, by construction.
Let $\delta_1 a$, $\delta_2 b$,  $\delta_3 c$, be the segments of
intersection of the three lines with $T$. 
\begin{figure}[htbp]
\centerline{\epsfxsize=3.5in \epsffile{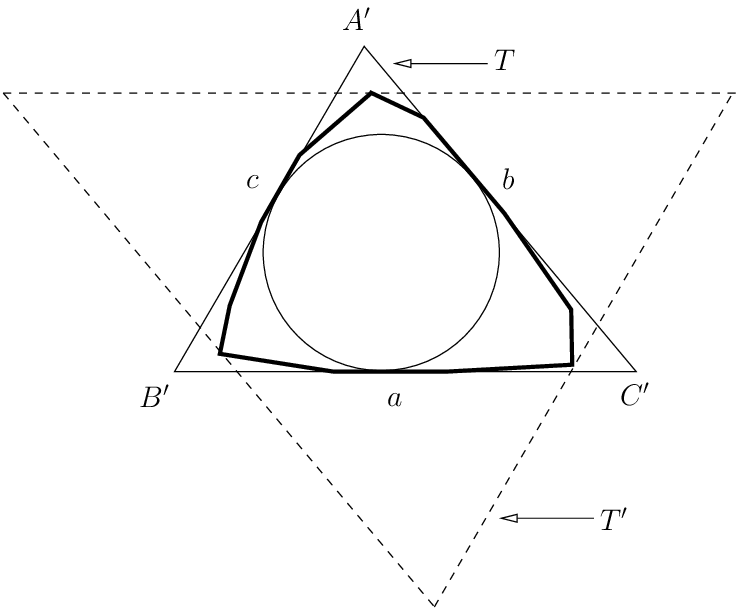}}
\caption{$P$ (in bold lines), $T$ (in solid lines) 
and $T'$ (in dashed lines).}
\label{f9}
\end{figure}

By the assumption of Case 2.2., we have
$ h'_\xi \leq \delta h_\xi$, for each $\xi \in \{a,b,c\}$. 
where $h'_a$, $h'_b$, $h'_c$ denote the altitudes from $A'$,
$B'$, and $C'$ in the three smaller similar triangles incident to
$A'$, $B'$, and $C'$. It follows that 
$$ \delta_1 + \delta_2 + \delta_3 = 
\frac{h'_a}{h_a} + \frac{h'_b}{h_b} + \frac{h'_c}{h_c} \leq 3\delta. $$

It is easily seen that the similarity ratio between $T'$ and $T$ is 
$2-\delta_1 - \delta_2 - \delta_3$. By the previous bound, this ratio
is at least $2-3\delta \geq 1$. 
Observe that $P$ is incident to the three sides of the acute triangle $T'$. 
It is well-known that the minimum-perimeter triangle inscribed in a
given acute triangle $\Delta$ (\ie, with a vertex incident to each
side of $\Delta$) is the \emph{orthic} triangle of
$\Delta$~\cite[Theorem~17]{Ka61}; or see~\cite[Ch.~5]{RT57}.   
The vertices of the orthic triangle are the feet of the altitudes of 
$\Delta$. It is also known that the semiperimeter of the orthic triangle
of an acute triangle with semiperimeter $s$, and sides $x$, $y$ and $z$
is equal to 
$$ \frac{4s (s-x)(s-y)(s-z)}{xyz}. $$
In particular, since $a+b+c=1$, the semiperimeter of the orthic
triangle of $T$ is 
$$ \frac{2(\frac12-a)(\frac12-b)(\frac12-c)}{abc}. $$
Since the similarity ratio between $T'$ and $T$ is at least
$2-3\delta$, by taking into account~\eqref{E9}, we obtain that
the semiperimeter of the orthic triangle of $T'$ is at least 
\begin{align*}
\frac{2(2-3\delta)(\frac12-a)(\frac12-b)(\frac12-c)}{abc}
&= 2\left(\frac{3}{\lambda}-7\right) \left(\frac{1}{2a}-1\right) 
\left(\frac{1}{2b}-1\right) \left(\frac{1}{2c}-1\right)\\
&\geq 2\left(\frac{3}{\lambda}-7\right) \left(\frac{1}{2\lambda}-1\right)^3. 
\end{align*}
Since $P$ is incident to the three sides of $T'$, its 
semiperimeter $p/2$ is bounded from below by the above expression, thus
\begin{equation} \label{E10}
\frac{p}{2} \geq 2\left(\frac{3}{\lambda}-7\right)
\left(\frac{1}{2\lambda}-1\right)^3.  
\end{equation}

We now bound from above the length of the third barrier $B_3$.
Recall that $a \geq b \geq c$. We have $\angle{C} \leq \pi/3$, 
thus $-\cos (\angle{C} +\pi/3) \leq 1/2$. 
One can deduce from~\cite[Section 5]{GP68} (or from~\cite[Ch.~6]{RT57})
and by using our assumptions in Case 2 that 
$$ |B_3|^2 = a^2 +b^2 -2ab \cos (\angle{C} +\pi/3)
\leq a^2 +b^2 +ab \leq 3 \lambda^2, $$
hence $|B_3| \leq \lambda \sqrt{3}$. 
Taking into account~\eqref{E10}, under the assumptions in Case 2.2,
the approximation ratio is at most 
\begin{equation}\label{E11}
\rho_2 := 
\frac{\lambda \sqrt{3}}{2\left(\frac{3}{\lambda}-7\right)
  \left(\frac{1}{2\lambda}-1\right)^3}. 
\end{equation}

Clearly, the approximation ratio of algorithm {\bf A2} is at most
$\rho = \max\{\rho_1, \rho_2\}$.
To balance Cases 1 and 2.1 with Case 2.2,
we let $\lambda$ be the solution to the equation
$\rho_1(\lambda) = \rho_2(\lambda)$ below; recall~\eqref{E8} and~\eqref{E11}:
\begin{equation}\label{rho}
1 + \frac{1}{\lambda(\pi + 2)}
=
\frac{\lambda \sqrt{3}}{2\left(\frac{3}{\lambda}-7\right)
  \left(\frac{1}{2\lambda}-1\right)^3}. 
\end{equation}
A routine calculation shows that
$\lambda = 0.3403\ldots$ and, correspondingly,
$\delta = 3 - 1/\lambda = 0.0615\ldots$ and $\rho_1 = \rho_2 = 1.5715\ldots$.
We conclude that the approximation ratio of 
algorithm {\bf A2} is at most $1.5716$, as claimed.

The largest circle inscribed in a convex polygon can be found
by linear programming in linear time~\cite{Me84}.
Computing $B_3$ given $T$ takes constant time, 
thus $B_3$ can be computed in $O(n)$ time.
Recall that $B_1$ and $B_2$ can be computed in $O(n)$ time too.
Consequently, the algorithm {\bf A2} takes $O(n)$ time. 
\end{proof}

It is easy to see that the connected barrier computed by  
{\bf A2} is not optimal in general (in the class of connected barriers).
The square gives an easy example.
The length of the third barrier from the left in Fig.~\ref{f1} is 
$1+\sqrt3$, while the length of the barrier computed by {\bf A2} 
is $3$ ($|B_1|=|B_2|=3$, $|B_3|=\infty$). This example shows a lower bound
of $1.098\ldots$ on the approximation ratio of the algorithm {\bf A2}.

\section{Single-arc barriers} \label{sec:single-arc}

Since algorithm {\bf A1} computes a single-arc barrier, and we have 
$\opt_{\rm conn}(P) \leq \opt_{\rm arc}(P)$, we immediately 
get an approximation algorithm with ratio 
$\frac{\pi+5}{\pi+2} = 1.5834\ldots$ for computing single-arc barriers. 

\begin{theorem} \label{T7}
Given a convex polygon $P$ with $n$ vertices, a single-arc polygonal barrier
whose length is at most $\frac{\pi+5}{\pi+2} = 1.5834\ldots$ times
longer than the optimal can be computed in $O(n)$ time.
\end{theorem} 

One may ask whether the single arc barrier computed by  
{\bf A1} is optimal (in the class of single arc barriers).
We show that this is not the case:
Consider (a sufficiently fine polygonal approximation of) 
a Reuleaux triangle $T$ of (constant) width 1, with three
vertices $a$, $b$, $c$. Now slightly shave the two corners at $b$ and
$c$ and obtain a convex body $T'$ of (minimum) width $1-\eps$
along $bc$. The algorithm {\bf A1} would return a curve of length close to 
$\pi/2 + 1 = 2.57\ldots$, while the optimal curve has length at most
$2\pi/3 + 2 - \sqrt{3} = 2.36\ldots$. 
This example shows a lower bound
of $1.088\ldots$ on the approximation ratio of the algorithm {\bf A1}. 
On the other hand, we believe that the approximation ratio of {\bf A1}
is much closer to this lower bound than to $1.5834\ldots$. 

We next present an improved version {\bf A3} of our algorithm {\bf A1} that
computes the shortest single-arc barrier of the form shown in Fig.~\ref{f4}. 
Let $P$ be a convex polygon with $n$ sides, and let $\ell$ be a line
tangent to the polygon, \ie, $P \cap \ell$ consists of a vertex of
$P$ or a side of $P$. For simplicity assume that $\ell$ is the $x$-axis,
and that $P$ lies in the closed halfplane $y \geq 0$ above $\ell$. 
Let $T=(\ell_1,\ell_2)$ be a minimal vertical strip enclosing $P$. Let
$p_1 \in \ell_1 \cap P$ and $p_2 \in \ell_2 \cap P$, 
be the two points of $P$ of minimum $y$-coordinates on the two
vertical lines defining the strip. Let $q_1 \in \ell_1$ and $q_2 \in \ell_2$  
be the projections of $p_1$ and $p_2$, respectively, on
$\ell$, and $\arc(p_1,p_2) \subset \partial P$ be the polygonal arc
connecting $p_1$ and $p_2$ on the top boundary of $P$ .  

The $U$-curve corresponding to $P$ and $\ell$, denoted $U(P,\ell)$ is
the polygonal curve obtained by concatenating $q_1 p_1$,
$\arc(p_1,p_2)$, and $p_2 q_2$, in this order. Obviously, for any line
$\ell$, the curve $U(P,\ell)$ is a single-arc barrier for $P$. 
Let $U_{\min}(P)$ be the $U$-curve of minimum length over all directions
$\alpha \in [0,\pi)$ (\ie, lines $\ell$ of direction $\alpha$). 

We next show that given $P$, the curve $U_{\min}(P)$ can be computed
in $O(n)$ time.  
The algorithm {\bf A3} is very simple: instead of
rotating a line $\ell$ around $P$, we fix $\ell$ to be horizontal, and
rotate $P$ over $\ell$ by one full rotation (of angle $2\pi$). 
We only compute the lengths of the $U$-curves corresponding to lines
$\ell$, $\ell_1$, $\ell_2$, supporting one edge of the polygon. 
The $U$-curve of minimum length among these is output.
There are at most $3n$ such discrete angles (directions), and the
length of a $U$-curve for one such angle can be computed in constant
time from the length of the $U$-curve for the previous angle.
The algorithm is similar to the classic rotating calipers algorithm 
of Toussaint~\cite{T83}, and it takes  $O(n)$ time by the previous
observation. 

To justify its correctness, it suffices to show that if each of the
lines $\ell$, $\ell_1$, $\ell_2$ is incident to only one vertex of
$P$, then the corresponding $U$-curve is not minimal. 

\begin{lemma}\label{L11}
Let $P$ be a convex polygon tangent to a line $\ell$ at a vertex $v
\in P$ only, and tangent to $\ell_1$ and $\ell_2$ at vertices $p_1$ and $p_2$
only. Then the corresponding $U$-curve $U(P,\ell)$ is not minimal.
\end{lemma}
\begin{proof}
For convenience, assume that $\ell$ is horizontal, and that $P$ lies
in the closed halfplane above $\ell$. Refer to Fig.~\ref{f7}.
\begin{figure}[htbp]
\centerline{\epsfxsize=2.5in \epsffile{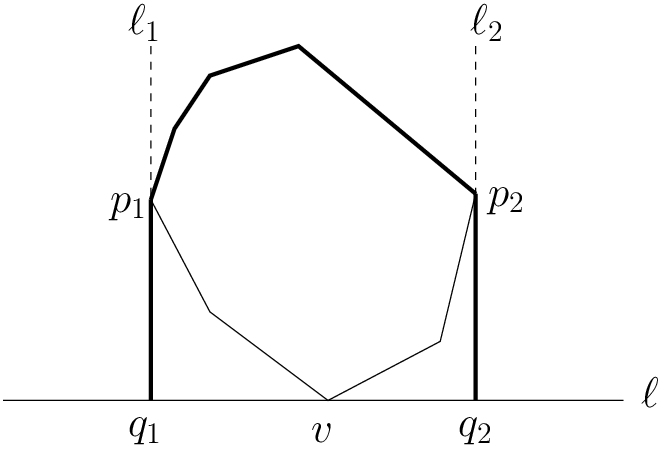}}
\caption{The curve $U(P,\ell)$.}
\label{f7}
\end{figure}

Let $p_1,q_1 \in \ell_1$ and $p_2,q_2 \in \ell_2$ be as defined
earlier. Observe that $v$ belongs to the closed segment $q_1 q_2$.
If $v=q_1$ (hence $v=q_1=p_1$) or $v=q_2$ (hence $v=q_2=p_2$),
then by rotating $P$ (clockwise or counterclockwise, as needed) around 
$v$ by a small angle, the length of the curve $U(P,\ell)$
decreases. So we can assume that $v$ lies in the interior of the
segment $q_1 q_2$. Observe that if $P$ rotates clockwise or
counterclockwise by a small angle around $v$, $p_1$ and $p_2$ remain
the same, so the angle $\angle{p_1 v p_2}$ also remains the same.  
Put $\alpha= \angle{q_1 v p_1}$,  $\beta= \angle{p_1 v p_2}$, 
and  $\gamma= \angle{p_2 v q_2}$, so $\alpha+ \beta+ \gamma=\pi$. 
Put $a= |p_1 v|$, $b=|p_1 p_2|$, and $c=|v p_2|$. The length of 
$U(P,\ell)$ for this angle $\alpha$ is
$$ f(\alpha)= a \sin \alpha + |\arc(p_1, p_2)| + c \sin \gamma. $$

The first two derivatives of $f(\cdot)$ are
\begin{align*}
f'(\alpha) &= a \cos \alpha - c \cos (\pi-\alpha-\beta)= 
a \cos \alpha - c \cos \gamma. \\
f''(\alpha) &= -a \sin \alpha - c \sin (\pi-\alpha-\beta)=
-a \sin \alpha - c \sin \gamma. 
\end{align*}

Since $\alpha, \gamma \in (0,\pi/2)$, we have $f''(\alpha) <0$,
which means that $f(\alpha)$ is not a local minimum.
\end{proof}

\begin{theorem} \label{T2}
Given a convex polygon~$P$ with $n$ vertices, the single-arc barrier
(polygonal curve) $U_{\min}(P)$ can be computed in $O(n)$ time. 
\end{theorem} 

Obviously, the approximation ratio of the algorithm {\bf A3}
is not worse than that achieved by algorithm {\bf A1}, hence it
is also bounded by $\frac{\pi + 5}{\pi + 2} = 1.5834\ldots$.
One may ask again whether the single arc barrier computed by  
{\bf A3} is optimal (in the class of single arc barriers).
We show again that this is not the case. Consider the pentagon
with vertices 
$(0,\eps)$, $(-3,0)$, $(-1,-\eps)$, $(1,-\eps)$, $(3,0)$. 
The optimal curve is no longer than the curve
$((-3,0),(-1,-\eps),(0,\eps),(1,-\eps),(3,0))$, 
whose length is 
$6 + O(\eps^2)$. On the other hand, the algorithm {\bf A3} returns a
curve of length $6 + \Omega(\eps)$. Let now $\eps$ be sufficiently small.

\begin{figure} [hbtp]
\centerline{\epsfxsize=3.3in \epsffile{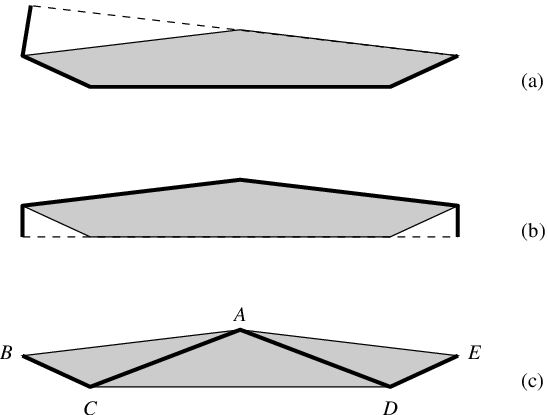}}
\caption{A pentagon with five vertices
$A = (0,h)$,
$B = (-x,y)$,
$C = (-1,0)$,
$D = (1,0)$,
$E = (x,y)$,
where $x = 1.4507\ldots$, $y = 0.2072\ldots$, and $h = 0.3806\ldots$.
The algorithm {\bf A3} returns a barrier of length $3.3364\ldots$
as in (a) or (b), but the barrier in (c) has a shorter length of $3.132\ldots$.
This gives a lower bound of $1.065\ldots$ on the approximation ratio
of the algorithm.}
\label{pentagon}
\end{figure}

We can fine-tune (numerically) the above pentagon to obtain a lower
bound of $1.065\ldots$ on the approximation ratio of {\bf A3}. See
Fig.~\ref{pentagon}.

\section{Arbitrary barriers} \label{sec:arbitrary}

\begin{theorem} \label{T3}
Given a convex polygon $P$ with $n$ vertices, a (possibly disconnected)
barrier for $P$, whose length is at most $\frac{1}{2}+ \frac{2
+\sqrt{2}}{\pi}=1.5867\ldots$ times longer than the optimal
can be computed in $O(n)$ time.
\end{theorem} 
\begin{proof}
Consider the following algorithm {\bf A4} which computes a 
(generally disconnected) barrier.
First compute a minimum-perimeter rectangle $R$ containing $P$;
refer to Fig.~\ref{f6}.
Let $a$,$b$,$c$,$d$,$e$,$f$,$g$,$h$, \linebreak
$i$,$j$,$k$,$l$ be the $12$
segments on the boundary of $R$ as shown in the figure; here $b$, $e$,
$h$ and $k$ are (possibly degenerate) segments on the boundary of $P$
contained in the left, bottom, right and top side of $R$. 
Let $P_i$, $i=1,2,3,4$  be the four polygonal paths on $P$'s boundary,
connecting these four segments as shown in the figure.
\begin{figure}[htbp]
\centerline{\epsfxsize=5in \epsffile{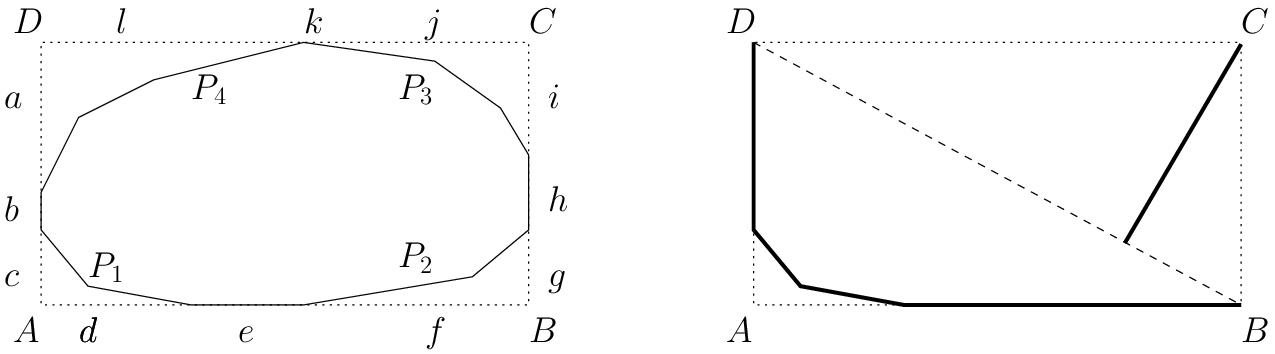}}
\caption{The approximation algorithm {\bf A4}.}
\label{f6}
\end{figure}

Consider four barriers for $P$, denoted $B_i$, for $i=1,2,3,4$.
$B_i$ consists of the polygonal path $P_i$ extended at both ends
on the corresponding rectangle sides, and the height from the opposite
rectangle vertex in the complementary right-angled triangle; see
Fig.~\ref{f6}~(right). The algorithm returns the shortest of
the four barriers. Let $h_A$, $h_B$, $h_C$, $h_D$ denote the four
altitudes from $A$, $B$, $C$, and $D$, respectively, in the
right-angled triangles $\Delta{ABD}$, $\Delta{BCA}$, 
$\Delta{CBD}$, and $\Delta{DAC}$. 
We have $|h_A|=|h_B|=|h_C|=|h_D|$ and the following other equalities:  
\begin{align*}
|B_1| &= |a| + |b| +|P_1| + |e| + |f| + |h_C|,\\
|B_2| &= |d| + |e| +|P_2| + |h| + |i| + |h_D|, \\
|B_3| &= |g| + |h| +|P_3| + |k| + |l| + |h_A|, \\
|B_4| &= |j| + |k| +|P_4| + |b| + |c| + |h_B|.
\end{align*}
By adding them up yields
\begin{align} \label{E25}
\sum_{i=1}^4 |B_i| &= \big(|b| + |e| + |h| + |k| + \sum_{i=1}^4 |P_i|\big) +
\big(|a| + \ldots + |l|\big) + \big(|h_A| + |h_B| + |h_C| + |h_D|\big)  \nonumber \\
&= \per(P) + \per(R) +  4|h_A| .
\end{align}

The length of the altitude $|h_A|$ in the right-angled triangle
$\Delta{ABD}$ is given by the formula 
$$ |h_A|= \frac{xy}{\sqrt{x^2 + y^2}}, $$
where $x$ and $y$ are the lengths of the two sides of $R$.
By Lemma~\ref{L10} we have
$$ \per(R) =2(x+y) \leq \frac{4}{\pi} \, \per(P). $$
Under this constraint, $|h_A|$ is maximized for
$x=y=\frac{\per(P)}{\pi} $, namely

$$ |h_A| \leq \frac{\per(P)}{\pi \sqrt{2}} 
\quad \Rightarrow \quad 4|h_A| \leq 
\frac{2 \sqrt{2}}{\pi}  \, \per(P) . $$
Hence~\eqref{E25} yields
$$ \min_{i} |B_i| \leq \frac{1}{4}
\left( 1+ \frac{4}{\pi} + \frac{2 \sqrt{2}}{\pi} \right) \per(P) . $$

Recall that $\per(P)/2$ is a lower bound on the length of
an optimal solution. The ratio between the length of the solution and the
lower bound on the optimal solution is
$$ \frac{\pi +4 + 2 \sqrt{2} }{2 \pi} =
\frac{1}{2}+ \frac{2 +\sqrt{2}}{\pi}=1.5867\ldots $$
Consequently, the approximation ratio of the algorithm {\bf A4} is
$ \frac{1}{2}+ \frac{2 +\sqrt{2}}{\pi}=1.5867\ldots $.
The algorithm takes $O(n)$ time, since computing the
minimum-perimeter rectangle containing $P$ takes $O(n)$ time
with the standard technique of rotating calipers~\cite{PS85,T83}.
This completes the proof of Theorem~\ref{T3}.
\end{proof}

The above analysis of the approximation ratio of {\bf A4} is tight
for a circle, in particular if $P$ is a regular $n$-gon with $n$
tending to infinity.
Indeed, for a unit-radius circle, $\per(P)/2=\pi$
while the length of the barrier computed by {\bf A4} is
$2+\pi/2+\sqrt{2}$.
The approximation ratio is exactly $(2+\pi/2+\sqrt{2})/\pi=1.5867\ldots$
in this case.

\section{Interior-restricted versus unrestricted barriers}
\label{sec:interior} 

In certain instances, it is infeasible to construct barriers guarding
a specific domain outside the domain (which presumably belongs to
someone else). We call such barriers constrained to the interior and
the boundary of the domain, \emph{interior-restricted}, or just
\emph{interior}, and all others \emph{unrestricted}. 
For example, all four barriers for the unit square illustrated 
in Fig.~\ref{f1} are interior barriers.

In the late 1980s,
Akman~\cite{A87} soon followed by Dublish~\cite{Du88} had reported
algorithms for computing a minimum interior-restricted barrier of a
given convex polygon (they refer to such a barrier as an \emph{opaque
minimal forest} of the polygon). Both algorithms however have been
shown to be incorrect by Shermer~\cite{Sh91} in 1991. 
He also proposed (conjectured) a new exact algorithm instead, but
apparently, so far no one succeeded to prove its correctness. 
To the best of our knowledge, the computational complexity of
computing a shortest barrier (either interior-restricted or
unrestricted) for a given convex polygon remains open.  

\bigskip
Next we show that a minimum \emph{connected interior} barrier for a
convex polygon can be computed efficiently:
\begin{theorem} \label{T4}
Given a convex polygon $P$, a minimum Steiner tree of the vertices of $P$
forms a minimum connected interior barrier for $P$.
Consequently, there is a fully polynomial-time approximation scheme for finding
a minimum connected interior barrier for a convex polygon.
\end{theorem}
\begin{proof}
Let $B$ be an optimal barrier. For each vertex $v \in P$, consider a
line $\ell_v$ tangent to $P$ at $v$, such that $P \cap \ell_v =\{v\}$.
Since $B$ lies in $P$, $\ell_v$ can be only blocked by $v$, so $v \in B$.
Now since $B$ is connected and includes all vertices of $P$, its
length is at least that of a minimum Steiner tree of $P$, as claimed.
Recall that the minimum Steiner tree problem for $n$ points in the
plane in convex position admits a fully polynomial-time approximation
scheme that achieves an approximation ratio of $1+\eps$ and runs in time
$O(n^6/\eps^4)$ for any $\eps > 0$~\cite{Pr88}.
\end{proof}

A minimum \emph{single-arc interior} barrier for a convex polygon can
be also computed efficiently. 
As it turns out, this problem is equivalent to that of finding a
shortest traveling salesman path (\ie, Hamiltonian path) for the $n$
vertices of the polygon.  

\begin{theorem} \label{T5}
Given a convex polygon $P$, a minimum Hamiltonian path of the vertices of $P$
forms a minimum single-arc interior barrier for $P$.
Consequently, there is an $O(n^2)$-time exact algorithm for finding
a minimum single-arc interior barrier for a convex polygon with $n$ vertices.
\end{theorem}
\begin{proof}
The same argument as in the proof of Theorem~\ref{T4} shows that
any interior barrier for $P$ must include all vertices of $P$.
By the triangle inequality,
the optimal single-arc barrier visits each vertex exactly once.
Thus a minimum Hamiltonian path of the vertices forms
a minimum single-arc interior barrier.

We now present a dynamic programming algorithm
for finding a minimum Hamiltonian path of the vertices of a convex polygon.
Let $\{v_0,\ldots, v_{n-1}\}$ be the $n$ vertices of the convex polygon
in counter-clockwise order;
for convenience, the indices are modulo $n$, \eg, $v_n = v_0$.
Denote by $\dist(i,j)$ the Euclidean distance between
the two vertices $v_i$ and $v_j$.
For the subset of vertices from $v_i$ to $v_j$
counter-clockwise along the polygon,
denote by $S(i,j)$ the minimum length of a Hamiltonian path
starting at $v_i$,
and denote by $T(i,j)$ the minimum length of a Hamiltonian path
starting at $v_j$.
Note that a minimum Hamiltonian path must not intersect itself.
Thus the two tables $S$ and $T$ can be computed by dynamic programming
with the base cases
$$
S(i, i+1) = T(i, i+1) = \dist(i, i+1)
$$
and with the recurrences
\begin{align*}
S(i, j) &= \min\{
	\dist(i, i+1) + S(i+1, j),
\,
	\dist(i, j) + T(i+1, j)
\},
\\
T(i, j) &= \min\{
	\dist(j, j-1) + T(i, j-1),
\,
	\dist(j, i) + S(i, j-1)
\}.
\end{align*}
Then the minimum length of a Hamiltonian path on the $n$ vertices is
$$
\min_i \min\{
\dist(i,i+1) + S(i+1, i-1),
\,
\dist(i,i-1) + T(i+1, i-1)
\}.
$$
The running time of the algorithm is clearly $O(n^2)$.
\end{proof}

\paragraph{Remark.} Observe that the unit square contains a disk of
radius $1/2$. According to the result of Eggleston mentioned earlier~\cite{E82},
the optimal (not necessarily interior-restricted) connected barrier
for a disk of radius $r$ has length $(\pi + 2)r$. This optimal barrier
is a single curve consisting of half the disk perimeter and two segments
of length equal to the disk radius. It follows that the optimal 
(not necessarily interior-restricted) connected barrier for the unit square
has length at least $(\pi + 2)/2=\pi/2+1=2.5707\ldots$. Compare this 
with the current best construction (illustrated in Fig.~\ref{f1},
third from the left) of length $1 +\sqrt{3}= 2.7320\ldots$.
Note that this third construction in Fig.~\ref{f1} gives the optimal connected
interior barrier for the square because of Theorem~\ref{T4}.
Further note that the first construction in Fig.~\ref{f1} gives the
optimal single-arc interior barrier because of Theorem~\ref{T5}.

\section{Conclusion}

Interesting questions remain open regarding the structure of optimal
barriers and the computational complexity of computing such barriers.
For instance:

\begin{itemize}
\itemsep 0.01in
\item [(1)] Does there exist an absolute constant $c \geq 0$ 
(perhaps zero) such that the following holds? The shortest barrier for
any convex polygon with $n$ vertices is a barrier consisting
of at most $n+c$ segments. 
\item [(2)] Is there a polynomial-time algorithm for computing 
a shortest barrier for a given convex polygon with $n$ vertices?
\item [(3)] Can one give a characterization of the class of convex
polygons whose optimal barriers are interior?
\end{itemize}

In connection with question (2) above, let us notice that the problem of
deciding whether a given barrier $B$ is an opaque set 
for a given convex polygon is solvable in polynomial time:

\begin{theorem} \label{T6}
Given a convex polygon $P$ with $n$ vertices, and a barrier $B$
with $k$ segments, there is a polynomial-time algorithm for deciding
whether $B$ is an opaque set for $P$.
\end{theorem}
\begin{proof}
Let $V(B)$ denote the $2k$ endpoints of the segments in $B$.
Consider the set of lines (directions) $\L$ determined either by pairs of
distinct points in $V(B)$ or that are incident to a point in 
$V(B)$ and tangent to $P$. Observe that $\L$ has $O(k^2)$ elements, 
and it can be easily constructed in $O(nk+k^2)$ time.  
If $B$ is not an opaque set for $P$, there exists a 
line $\ell \in \L$ such that a small rotation (clockwise or counterclockwise) 
around $\ell$ yields a direction, say $\ell^+$ or $\ell^-$, 
such that the projection of $B$ onto the line orthogonal to it
does not cover the projection of $P$ onto the same line.
That is, the union of the projection segments does not include
the segment which represents the projection of $P$. 
We say that the opaqueness condition fails with respect to 
$\ell^+$ or $\ell^-$.

To see this, take a line that intersects $P$ without intersecting
$B$. Fix a point $p$ in $P$ (in the interior or on the boundary of
$P$) on this line and rotate the line around $p$ until it hits a segment in $B$,
say at its endpoint $q$. Start rotating the line around $q$ until
either it becomes tangent to $P$ (as it leaves $P$), or 
it hits another segment endpoint in $V(B)$.

For a given $\ell \in \L$ the opaqueness condition for 
$\ell^+$ and $\ell^-$ can be easily checked in $O(n+k)$ time. 
Since there are $O(k^2)$ lines in $\L$, 
the overall opaqueness can be checked in $O((n+k) \cdot k^2)$ time.
Hence whether $B$ is an opaque set for $P$ can be determined
in $O(nk+k^2+(n+k) \cdot k^2)= O((n+k) \cdot k^2)$ time.
(A faster algorithm can be obtained by using rotational 
sweep~\cite[p.~328]{BCKO08}.)
\end{proof}

We have presented several approximation and exact algorithms for
computing shortest barriers of various kinds, for a given convex
polygon. The two approximation algorithms with ratios close to $1.58$
probably cannot be improved substantially without either increasing
their computational complexity or finding a better lower bound
on the optimal solution than that given by Lemma~\ref{L2}.
The question of finding a better lower bound is particularly intriguing,
since even for the simplest polygons, such as a square, we don't
possess any better tool. 
While much research up to date focused on upper or lower bounds for
specific example shapes, obtaining a polynomial time approximation
scheme (in the class of arbitrary barriers) for an arbitrary convex
polygon is perhaps not out of reach.

\end{document}